\documentclass[a4paper,UKenglish,cleveref,autoref,thm-restate]{lipics-v2021}

\newif\ifcameraready
\camerareadytrue

\usepackage[normalem]{ulem}
\usepackage{mdframed}
\usepackage{booktabs}
\usepackage{cleveref}
\usepackage{xspace}

\newcommand{\tx}{\mathsf{T}}
\newcommand{\obj}{\mathsf{obj}}
\newcommand{\readop}{\operatorname{read}}
\newcommand{\writeop}{\operatorname{write}}
\newcommand{\readset}{\texttt{read}}
\newcommand{\writeset}{\texttt{write}}

\newcommand{\acc}{\texttt{0xacc}\xspace}
\newcommand{\clock}{\texttt{0x6}\xspace}
\newcommand{\rand}{\texttt{0x8}\xspace}

\newmdenv[
  backgroundcolor=black!6,
  linecolor=black!25,
  linewidth=0.4pt,
  leftmargin=0pt, rightmargin=0pt,
  innerleftmargin=8pt, innerrightmargin=8pt,
  innertopmargin=6pt, innerbottommargin=6pt,
  skipabove=8pt, skipbelow=8pt,
]{findingbox}
\newenvironment{finding}[1][]{\begin{findingbox}\noindent\textbf{Finding\if\relax\detokenize{#1}\relax\else~(#1)\fi.}\enspace}{\end{findingbox}}

\lstdefinelanguage{toml}{
  morecomment=[l]{\#},
  morestring=[b]",
  moredelim=[s][\color{gray}]{[}{]},
  morekeywords={git,rev,package},
}

\title{Not All Reads Are Conflicts: A Write-Only Analysis of the Sui Blockchain}

\ifcameraready
\author{Haygen Tsoi}{University College London (UCL), United Kingdom}{siu.tsoi.24@alumni.ucl.ac.uk}{https://orcid.org/0009-0001-1840-9133}{}
\author{Alberto Sonnino}{Mysten Labs and University College London (UCL), United Kingdom}{alberto@mystenlabs.com}{https://orcid.org/0000-0001-5337-4741}{}
\author{Philipp Jovanovic}{Mysten Labs and University College London (UCL), United Kingdom}{p.jovanovic@ucl.ac.uk}{https://orcid.org/0000-0002-2202-1879}{}

\authorrunning{H. Tsoi, A. Sonnino, P. Jovanovic}
\Copyright{Haygen Tsoi, Alberto Sonnino, Philipp Jovanovic}
\fi


\ifcameraready
\nolinenumbers
\acknowledgements{We would like to thank Mysten Labs for their generous support and funding of this research.
}
\fi

\relatedversion{} 

\relatedversiondetails[
    linktext={arXiv preprint}
]{Full Version}
{https://arxiv.org/abs/2607.26691}

\supplement{}

\supplementdetails[
    subcategory={Source Code},
    linktext={Rust/Diesel Indexer and Analysis Scripts}
]{Software}
{https://github.com/krdecade-0/sui_analysis_code}

\funding{This work is partially funded by Mysten Labs}

\EventEditors{Aggelos Kiayias and Maria Kyropoulou}
\EventNoEds{2}
\EventLongTitle{8th Conference on Advances in Financial Technologies (AFT 2026)}
\EventShortTitle{AFT 2026}
\EventAcronym{AFT}
\EventYear{2026}
\EventDate{October 6--9, 2026}
\EventLocation{London, UK}
\EventLogo{}
\SeriesVolume{395}
\ArticleNo{20}

\begin{document}

\maketitle

\begin{abstract}
    Sui's object-centric data model enables parallel transaction execution, but realised performance is fundamentally bounded by workload contention. Prior empirical analyses of Sui have relied on ``read+write'' (R+W) conflict graphs inherited from account-based blockchains. Because Sui's engine serialises only on mutable shared access, R+W graphs contain spurious edges, bounding contention from above. In this paper, we adopt a complementary ``write-set-only'' (W-only) model in which every edge represents a real write-serialisation event, providing a lower bound on contention. Together, the two models bracket Sui's true execution-dependency structure.
Applying the W-only analysis to Sui mainnet data through 2025 yields three primary findings. First, removing read-only dependencies (notably the system clock) causes previously reported ``hub-and-spoke'' structures to collapse. The remaining contention topology is highly assortative and clique-dominated, with the W-only bound shaving roughly $30$--$40\%$ off the R+W estimate of Sui's optimal-parallelism headroom. Second, via union-find object grouping, we isolate DeepBook (Sui's native central limit order book). While it dominates contention by volume, its underlying logic does not impose disproportionate sequential bottlenecks. Finally, we quantify the economic cost of contention, showing that $10$--$50\%$ of the network's USD-denominated value flows through sequentially constrained execution paths, exposing it to potential ordering effects.

\end{abstract}

\section{Introduction}

Since the launch of Bitcoin in 2008~\cite{nakamoto2008bitcoin}, blockchain research has focused on scaling transaction throughput to compete with traditional payment processors like VISA. While early smart-contract platforms like Ethereum~\cite{ethereum} introduced expressive programmability, they have struggled with the ``ordering curse'': a sequential execution model where transactions must be interleaved through a single global state.
Sui was introduced as a high-performance alternative designed specifically for parallel execution. By adopting an object-centric data model, Sui enables parallel execution for transactions that touch disjoint sets of state. However, as with any parallel system, the actual performance realised in production is limited by the structure of the workload and the frequency of state contention~\cite{biton2025analysis}.

In this paper, we evaluate the residual parallelism available in Sui by analysing mainnet transaction data through the end of 2025. Prior empirical work~\cite{biton2025analysis} adopts a ``read+write'' (R+W) conflict model inherited from account-based blockchains, where a graph-based approach is used to measure contention. An edge is drawn between two transactions (nodes) if they touch the same object and at least one modifies it. Thus, more edges in the graph mean more contention, which is equivalent to more serialisation and less parallelism. We instead adopt a different ``write-only'' (W-only) model, where an edge is drawn only if both transactions attempt to modify the same object. Because Sui's execution engine serialises only on mutable shared accesses (\Cref{sec:why_write_only}), a R+W graph contains edges the engine never has to honour (e.g., every user transaction's read of the system clock is counted as a conflict with the system writer), so it bounds contention from \emph{above}. By contrast, every edge in the W-only graph corresponds to a real write-serialisation event, so W-only bounds contention from \emph{below} (and parallelism from above). The two models bracket Sui's true execution-dependency structure.

Our investigation is guided by four core research questions:
\begin{enumerate}
    \item \textbf{RQ1:} Under a write-only conflict model, what structural topology does Sui's conflict graph exhibit, and how does it differ from the topology reported under R+W? (\Cref{sec:results})
    \item \textbf{RQ2:} How does the W-only estimate of achievable parallelism compare quantitatively to the R+W estimate at matched load regimes? (\Cref{sec:results})
    \item \textbf{RQ3:} Is write contention concentrated in a small number of smart-contract applications (e.g. DeepBook), or is it widely spread? (\Cref{sec:contention})
    \item \textbf{RQ4:} What fraction of Sui's economic throughput (value transferred) flows through sequential-burden transactions vs. parallelisable ones? (\Cref{sec:value})
\end{enumerate}

To address these questions, we apply the W-only model to Sui mainnet transaction data through the end of 2025. The W-only graph reveals a highly assortative, clique-dominated topology, fundamentally different from the disassortative, star-like structures observed under R+W. The shift is driven by the collapse of read-side dependencies on the system clock: the edges that previously connected every user transaction to the clock disappear, leaving only a small system-to-system clique. Using the same LSP/$\chi$ metric as Biton \& Friedman~\cite{biton2025analysis} and matching their load regimes, our W-only estimate of achievable parallelism is roughly $30$--$40\%$ tighter at routine load ($\approx 1\times$ vs.\ their $1.2$--$1.4\times$) and trims their high-load ceiling from $\approx 5\times$ toward $\approx 3\times$. At the application level, a union-find grouping over write co-occurrence identifies the DeepBook ecosystem (a giant component containing $87.3\%$ of all application-level conflicts), and shows that DeepBook contributes to contention strictly by transaction volume rather than by uniquely contention-inducing logic, a surprising outcome for a single shared resource like a central limit order book. Finally, mapping graph structures to USD-denominated balance flow shows that $10$--$50\%$ of network value remains gated by sequential execution paths (with the remaining $50$--$90\%$ flowing through parallel paths), bounding the share of value that is potentially currently exposed to reordering-based MEV strategies.

Our primary contributions are as follows:
\begin{itemize}
    \item \textbf{Semantic re-evaluation of conflict modelling in Sui using a write-only conflict model.} We leverage a W-only conflict model whose edges correspond to guaranteed write-serialisation events that Sui's engine must preserve, and show that the R+W conflict model inherited from account-based systems systematically introduces non-execution dependencies under Sui's object-centric model. We propose a complementary W-only model aligned with mutable shared-object serialisation semantics.
    \item \textbf{Application-level partitioning via union-find.} We identify the DeepBook ecosystem in the conflict graph and compare the contention structure of DeepBook-touching versus non-DeepBook subgraphs.
    \item \textbf{Mapping graph structure to economic flow.} We map graph structures to USD-denominated balance flows, measuring the fraction of network value gated by sequential execution paths.
\end{itemize}
\section{Background}
\label{sec:background}

In this section, we present the background concepts and definitions that our analysis relies on. 
We first describe Sui's object-centric execution model, including the versioning rules that govern parallel mutability and serialisation, which together are the foundation of its parallelism (\Cref{sec:sui_model}). 
This motivates our choice of a write-only conflict model for measuring contention (\Cref{sec:why_write_only}). 
Finally, we define the graph metrics we use to quantify parallelism (\Cref{sec:graph_definitions}).

\subsection{Sui's Object-Centric Execution Model}
\label{sec:sui_model}

In Sui~\cite{sui}, every on-chain entity, such as a coin, a token, a smart contract's state, a deployed package, etc., is an \emph{object}. 
Each object carries a globally unique 256-bit identifier that is fixed for its lifetime, a version number, and an owner, and is a typed instance of a Move resource. 
Sui exposes three ownership classes that differ in how transactions may access them: \emph{address-owned} objects, reachable only through a single account; \emph{immutable} objects, frozen and readable by anyone; and \emph{shared} objects, mutable and reachable concurrently by any transaction. 
This object-level granularity is the foundation of Sui's parallelism: two transactions whose object sets are disjoint carry no execution dependency and may run on separate cores.
Whether two transactions that share an object must be serialised depends on how each of them accesses it. 
Sui coordinates this through a form of deterministic \emph{multi-version concurrency control} (MVCC) in which object versions are assigned by the consensus engine, Mysticeti~\cite{mysticeti}, rather than at execution time, and distinguishes three primary access modes:
\begin{enumerate}
    \item \textbf{Owned/Immutable:} These objects never trigger serialisation across transactions, since an owned object has a single writer by construction and an immutable object can only be read. The access pins to whichever version is current at consensus-sequencing time and does not advance the version counter.
    \item \textbf{Shared (Read-Only):} Many transactions can read the same shared object concurrently without blocking, and such reads leave the version counter untouched.
    \item \textbf{Shared (Mutable):} This is the only access that can induce an execution dependency. Two transactions that both write the same shared object must execute sequentially, and a mutating access consumes the object's current version $v$ and produces version $v{+}1$, advancing the counter.
\end{enumerate}

Because every version an access will read is fixed at consensus time, Sui needs no read locks.
Readers of an object $\obj$ at version $v$ (written $\obj{@}v$) are mutually independent, and a reader of $\obj{@}v$ is independent even of the writer that produces $\obj{@}v{+}1$, since the two touch different versions; the execution scheduler therefore treats an immutable access as a literal no-op.

User activity reaches the network as \emph{programmable transaction blocks} (PTBs): an atomically executed sequence of commands (e.g., Move calls, coin splits and merges, object transfers) that either commits in full or aborts with no effect~\cite{suidocs}. 
A PTB does not discover state at run time; it declares each input object up front, together with the version it will read.
Its read and write sets are thus fixed before execution and can be reconstructed statically from a transaction trace, the property on which our analysis rests.

A transaction's route through the network follows from the objects it declares.
One whose inputs are all owned or immutable takes the \emph{fast path} and is certified directly by validators, without consensus; one declaring any mutable shared object takes the \emph{consensus path} and is sequenced by Mysticeti against all competing transactions.
Because only the consensus path can serialise transactions of distinct users, it is the conflict structure of that path that the rest of the paper measures.

Executed transactions are finalised and recorded in \emph{checkpoints}.
Unlike a conventional block, a checkpoint is assembled \emph{after} execution: the checkpoint builder takes a contiguous run of consensus-ordered transactions, closes it under causal dependency, and certifies it with a quorum of validator signatures~\cite{biton2025analysis}.
Each checkpoint carries a monotonically increasing sequence number and is irreversible once certified.
Since the Mysticeti upgrade, Sui emits roughly four checkpoints per second, and one checkpoint may bundle several consensus commits, each contributing a system \emph{prologue} transaction that advances on-chain system state (\Cref{sec:why_write_only}).
A checkpoint is thus a finalised, deterministically ordered batch of transactions with fully known object accesses, the natural unit at which to study execution dependencies; following Biton \& Friedman~\cite{biton2025analysis}, we build one conflict graph per checkpoint.

\subsection{Graph Properties and Structures}
\label{sec:graph_definitions}

We use multiple graph properties to characterise our conflict graphs and determine potential parallelizability. These metrics are adapted from prior work~\cite{biton2025ethereum}.
\begin{itemize}
    \item \textbf{Density:} The ratio of existing edges to the maximum possible edges, measuring how closely a graph approaches a complete graph.
    \item \textbf{Diameter:} The longest shortest path between any two nodes in a connected graph.
    \item \textbf{Chromatic number ($\chi$):} The minimum number of colours needed to colour the nodes so that no two adjacent nodes share the same colour.
    \item \textbf{Clique number:} The number of nodes in the largest complete subgraph.
    \item \textbf{Longest simple path (LSP):} The length of the longest path that does not repeat any visited nodes.
    \item \textbf{Largest connected component (LCC):} The size of the connected subgraph with the highest number of nodes.
\end{itemize}

When parallel execution is not possible, Sui switches to sequential execution. The \textbf{LCC} provides the upper bound of sequential dependency for concurrently accessed transactions, while the \textbf{LSP} provides the lower bound of the sequential chain. We identify two primary structural archetypes:
\begin{itemize}
    \item \textbf{Clique-like:} A near-complete subgraph formed when multiple transactions write to the same hot object(s), creating dense conflict edges.
    \item \textbf{Hub-and-spoke (Star-like):} A centralised structure where a ``hub'' transaction writes to multiple hot objects, while ``spoke'' transactions write to a single hot object, connecting them to the hub but not to each other.
\end{itemize}

\subsection{Conflict Modelling: R+W vs. W-Only}
\label{sec:why_write_only}

Prior empirical analyses of Sui~\cite{biton2025analysis} adopt a \emph{read+write} (R+W) conflict model inherited from account-based blockchains such as Ethereum. Under this model, two transactions $\tx_1$ and $\tx_2$ conflict if they touch the same object and at least one of them writes.
Formally, two transactions $\tx_1$ and $\tx_2$ are connected by an edge in the R+W conflict graph iif
$$
(\writeset(\tx_1) \cap (\readset(\tx_2) \cup \writeset(\tx_2))) \cup
(\writeset(\tx_2) \cap (\readset(\tx_1) \cup \writeset(\tx_1))) \neq \emptyset
$$
where $\writeset(\tx_i)$ denotes the write set and $\readset(\tx_i)$ denotes the read set of transaction $\tx_i$.
Because Sui's scheduler does not serialise on shared-object reads, an R+W graph contains edges that the engine never has to honour, such as every read of the system clock \texttt{0x6}. 
R+W therefore provides an \emph{upper bound on contention}.

In this paper, we adopt a \emph{write-only} (W-only) conflict model as a complementary view: two transactions conflict only if they both write the same object.
Formally, two transactions $\tx_1$ and $\tx_2$ are connected by an edge in the W-only conflict graph iif
$$
\writeset(\tx_1) \cap \writeset(\tx_2) \neq \emptyset
$$
Every edge in our graph corresponds to a guaranteed write-serialisation event that Sui's engine must preserve under the mutable shared-object semantics, so W-only provides a \emph{lower bound on contention} and a \emph{practical upper-bound on achievable parallelism}.

\paragraph*{Advantages of the W-Only Model: No Artificial Read Dependencies}
Consider two transactions sequenced adjacent to each other in a single checkpoint, both touching a shared object $\obj$ at current version $v_0$:
\[
    \tx_{r_0}: \readop\, \obj{@}v_0
    \qquad\text{and}\qquad
    \tx_{w_1}: \writeop\, \obj\ (v_0 \to v_1).
\]
Under Sui's MVCC model, both transactions pin to $\obj{@}v_0$ as input. The scheduler emits no dependency from $\tx_{r_0}$'s read, and the barrier it emits for $\tx_{w_1}$'s write is only against \emph{other writers} of $\obj$. The two transactions execute in parallel. An R+W graph places an edge between them; a W-only graph does not.

W-only edges align directly with the four Sui subsystems that enforce serialisation, all of which key off the exclusive-access (\texttt{Mutable}) flag~\cite{sui_src_mutability_enum}:
\begin{itemize}
    \item \textbf{Execution scheduler:} \texttt{Mutable} access emits barrier dependencies; \texttt{NonExclusiveWrite} emits ordinary dependencies; \texttt{Immutable} access emits no edges at all~\cite{sui_src_scheduler}.
    \item \textbf{Shared-object version manager:} Only exclusively-accessed inputs advance the per-object version counter. Read-only accesses do not bump the version~\cite{sui_src_version_manager}.
    \item \textbf{Per-object congestion control:} Execution cost is accumulated only for exclusively-accessed inputs. Read-only accesses contribute zero cost and cannot cause deferral~\cite{sui_src_congestion_per_object}.
    \item \textbf{Checkpoint-level congestion accounting:} The filter explicitly drops read-only inputs and keeps only mutated objects for downstream processing~\cite{sui_src_congestion_checkpoint}.
\end{itemize}

By stripping read-side edges, the W-only model isolates application-driven contention from read-only hot system objects. The clearest example is the system clock (\clock). In a read+write model, almost every transaction in a checkpoint ``conflicts'' with the clock because they all read it for timestamps. This creates a massive, artificial ``star'' structure in the conflict graph. Each consensus commit emits a single \emph{prologue} transaction that mutates the clock, and Sui's checkpoint builder aggregates multiple consensus commits into a single checkpoint. A typical checkpoint therefore contains several prologue transactions, each writing \clock.

Under R+W, the resulting picture is a giant hub-and-spoke web: every prologue is a hub, and every user transaction in the checkpoint is a spoke, because all of them read the clock for timestamps. Under W-only, the spokes disappear, user reads do not produce edges, and what remains is a small clique of prologue-to-prologue writes. The character of the conflict is now completely different: the clock no longer appears to contend with user transactions, only with itself. This lets us cleanly separate clock activity from application contention and quantify it: in \Cref{sec:contention} we report that this system-to-system clique accounts for $68\%$ of W-only conflict events, isolating it as pure system housekeeping with no user-facing impact on parallelism.

The same logic applies to other per-commit system objects, such as the randomness state (\texttt{0x8}) and the accumulator root (\texttt{0xacc})~\cite{sui_src_system_singletons}, whose system-only writers form analogous prologue-style cliques when checkpoints aggregate multiple commits.

\paragraph*{Limits of the W-Only Model: Write-to-Read Forwarding}
There is one dependency class where the W-only model may under-count: write-then-read \emph{forwarding} within a single checkpoint. Consider the sequence:
\[
    \tx_{w_1}: \writeop\, \obj\ (v_0 \to v_1)
    \quad\Longrightarrow\quad
    \tx_r: \readop\, \obj{@}v_1
    \quad\Longrightarrow\quad
    \tx_{w_2}: \writeop\, \obj\ (v_1 \to v_2)
\]
The W-only graph captures the structural $\tx_{w_1}\!-\!\tx_{w_2}$ edge. However, the actual execution DAG may also contain $\tx_{w_1} \to \tx_r$, because $\tx_r$ pins to a version that only exists once $\tx_{w_1}$ finishes.
Whether this $\tx_{w_1} \to \tx_r$ edge actually constrains execution depends entirely on runtime timing. If $\tx_{w_1}$ completes before $\tx_r$ is scheduled, no wait occurs; $\tx_r$ consumes the materialized version and runs in parallel with $\tx_{w_2}$. If $\tx_{w_1}$ is slow, $\tx_r$ blocks. Because the static checkpoint trace cannot reveal this runtime timing, the W-only graph conservatively excludes these conditional edges. The W-only model is therefore best understood as the conservative core of guaranteed orderings: it overcounts no edges, though it may undercount these conditional forwarding chains. We treat W-only and R+W as complementary bounds throughout the remainder of the paper.

\section{Methodology}
\label{sec:methodology}

\subsection{Data Acquisition and Pipeline}
Transaction data was collected from the mainnet checkpoints into our PostgreSQL~\cite{PostgreSQL} database using a custom indexer written in Rust and Diesel~\cite{DieselORM}. Sui produces around 4 checkpoints per second, corresponding to approximately $345,600$ checkpoints per day. To collect a representative, long-term dataset across the history of the network without overwhelming our pipeline, we deterministically sampled every $3455^{th}$ checkpoint. 
At post-Mysticeti checkpoint production rates, this corresponds to approximately $100$ checkpoints per day. Earlier periods of the blockchain operated at a lower checkpoint frequency, around $\sim 1$~cp/s, resulting in fewer sampled checkpoints per day in the pre-Mysticeti era. This stride value was fixed globally rather than recalculated per day.

Once indexed, we extract the write-sets and construct the undirected conflict graphs for each checkpoint using NetworkX~\cite{NetworkX} in Python. The full database schemas and the SQL used to extract the write sets are reported in \Cref{app:schemas}.

\subsection{Quantifying Execution Parallelism}
To evaluate the potential for parallelism within Sui checkpoints, we employ two primary metrics adapted from prior works on Ethereum~\cite{biton2025ethereum, heimbach2023defi}.

First, we use the \emph{Longest Simple Path to Chromatic Number Ratio (LSP/$\chi$)}. This graph-based metric bounds the achievable parallel execution of a given checkpoint~\cite{optimising_blockchain}. The chromatic number ($\chi$) represents the theoretical minimum number of execution rounds required to process the graph without conflicts. The longest simple path (LSP) represents the largest unavoidable sequential dependency chain. Because finding the exact $\chi$, clique number, and LSP are NP-hard problems, we estimate them using heuristic algorithms: $\chi$ is approximated via NetworkX's greedy colouring algorithm, the clique number via a greedy heuristic, and the LSP via a heuristic depth-first search. While heuristic estimates may not give the exact values, our study focuses on relative comparison between R+W and W-only under identical estimation procedures, reducing systematic bias.

Second, we use \emph{Gas Usage} to weight these structural limits by computational cost. Gas usage represents the actual work done to execute transactions. By dividing the total gas used in a checkpoint by the gas consumed along the largest sequential path (the largest connected component or max clique), we estimate the potential speedup factor if all independent transactions were executed perfectly in parallel.
This gas-weighted variant is adapted from Heimbach et al.~\cite{heimbach2023defi}, who applied it to Ethereum. The count-based LSP/$\chi$ ratio used by prior work on Sui~\cite{biton2025analysis} treats every transaction equally: a $1$M-gas transaction contributes the same as a $1$K-gas one. This is appropriate for asking ``how many parallel rounds do we need?'' but misleading for ``how much wall-clock time can we save?''. Gas-weighting is more faithful to actual execution cost, and more conservative against long sequential chains composed of cheap transactions: a long chain of $1$K-gas transactions is correctly downweighted relative to a short chain of expensive ones.

\subsection{Application Isolation and Economic Modelling}
To answer our research questions regarding application-level contention and economic cost, we apply two specific modelling techniques to the base conflict graphs.

\paragraph*{Union-Find Object Grouping.}
\label{disjoint_object_set}
To isolate the impact of specific smart contracts (like DeepBook), we create union-find (disjoint set) groups over objects based on write co-occurrence. An object $\obj_1$ is related to $\obj_2$ if a transaction $\tx_1$ writes to both. This relationship is transitive; two objects $\obj_1$ and $\obj_2$ may end up in the same group via a chain of shared transactions, even if no single transaction touches both directly. This grouping provides an upper bound on the objects jointly involved in an application's execution ecosystem, which we use in \Cref{sec:contention} to partition the conflict graph.

\paragraph*{Balance-Change and Price Anchoring.}
\label{sec:value_methodology}
To measure the real-world economic value constrained by sequential execution, we analyse changes in coin balances using Sui's \texttt{derive\_balance\_changes} function. This function tracks value by subtracting custom fungible input coins, adding mutated/output coins, and tracking address balance changes from accumulator events. It focuses on fungible value, explicitly excluding NFTs or bespoke non-Coin objects.

By anchoring these balance changes to historical CoinMarketCap price data, we can calculate the USD-denominated value processed by any given transaction. We aggregate these values into two buckets: transactions that lie on the sequential burden (the Largest Connected Component) and those that can be executed in parallel.

\section{Conflict Graph Structure: Write-Only vs. Read+Write}
\label{sec:results}

The transition from a read+write (R+W) to a write-only (W-only) conflict model fundamentally alters the perceived structure of the Sui network. In this section, we characterise these structural shifts and explain why they yield a tighter bound on potential parallelism.

\subsection{Structural Topology: From Stars to Cliques}
\label{sec:0x6_collapse}

The most striking difference between our W-only results and the prior R+W analysis by Biton \& Friedman~\cite{biton2025analysis} is the shift in \emph{assortativity}. In a read+write model, Sui conflict graphs appear highly disassortative, characterised by ``stars'' where many transactions conflict with a single central hub.
As mentioned in \Cref{sec:background}, this artefact is explained almost entirely by the behaviour of system objects, such as the system clock (\clock). In an R+W analysis, the clock is a massive hub because virtually every transaction reads it for a timestamp, while only system transactions write to it. This creates a false-positive conflict edge between user transactions and the clock.

By utilising a W-only model, we remove these read-side edges, causing the artificial star structure to collapse. The user transactions disappear from the clock's component entirely. What remains are the true execution dependencies: multiple transactions attempting to modify the same application state.

Consequently, our W-only graphs are highly assortative. This suggests that actual execution contention on Sui is not a ``hub-and-spoke'' phenomenon, but rather a ``clique-like'' one. Transactions do not cluster around a single reader; they conflict in small, dense groups that are fundamentally unparallelizable.

\begin{figure}[t]
    \centering

    \includegraphics[width=\linewidth]{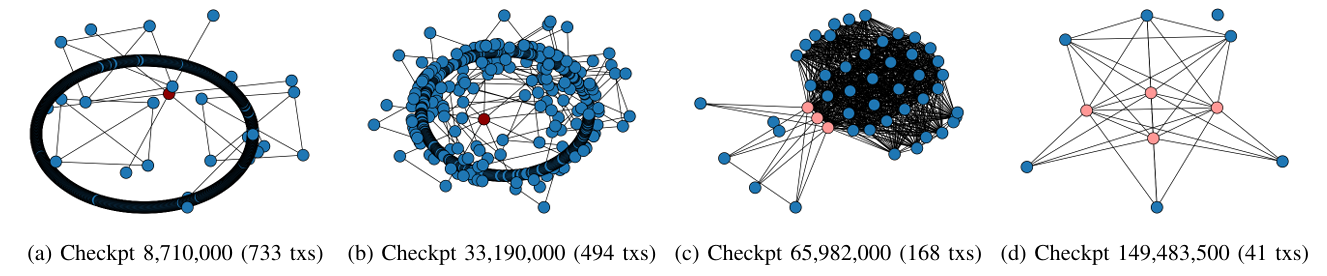}
    \caption{Biton \& Friedman's R+W checkpoint conflict graphs visualised~\cite{biton2025analysis}}

    \vspace{0.5em}

    \includegraphics[width=\linewidth]{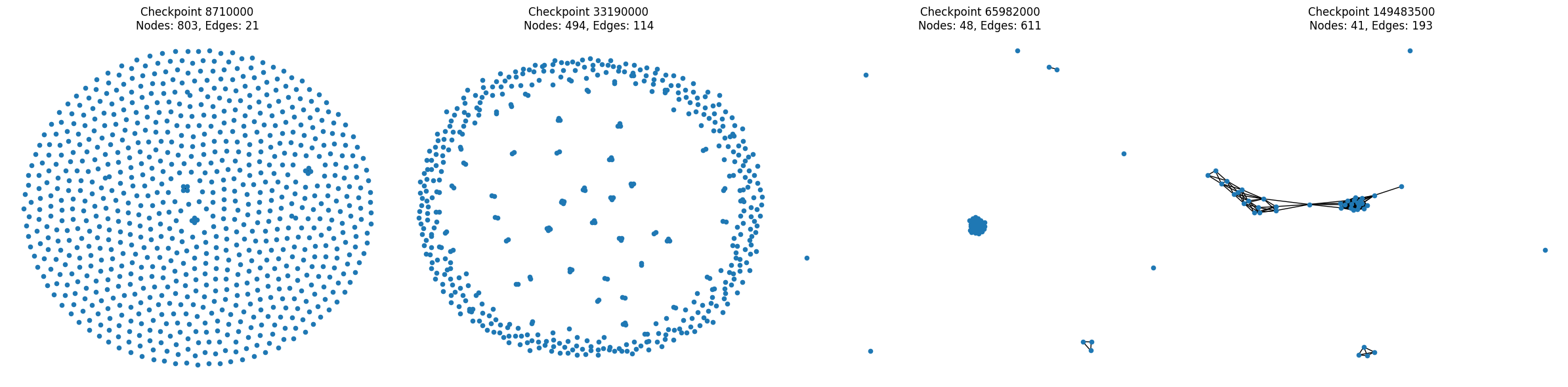}

    \caption{Our W-only conflict graphs visualised in the same checkpoints}
    \label{fig:read_write_checkpoints_vs_write_only_checkpoints}
\end{figure}

\Cref{fig:read_write_checkpoints_vs_write_only_checkpoints} shows the comparison of Biton \& Friedman's read+write conflict graphs with our write-only conflict graphs for the same checkpoints. The dense ``hub-and-spoke``'' structures observed in the read+write graphs largely disappear in our write-only graphs, indicating that much of the apparent contention originates from shared read access.

\subsection{Quantifying the Lower Bound}

Because we strip the non-execution read-edges, our measured contention is lower (fewer edges), but the resulting sequential chains are strictly execution-bound. This yields a sound lower bound on the contention Sui's scheduler must honour, and equivalently a (near-)upper bound on the speedup it could achieve.

\Cref{table:read_write_vs_write_only} compares the core graph metrics between the two models. Our W-only conflict graphs exhibit lower median density, clique number, and Largest Connected Component (LCC). Crucially, the topology shifts from disassortative (high-degree nodes connected mostly to low-degree ones, the signature of hub-and-spoke stars) to highly assortative (high-degree nodes connected to each other in clique-like clusters). The corresponding degree-assortativity coefficient jumps from $-0.500$ to $\approx+1$, quantifying the elimination of the read-induced stars.

\begin{table}[t]
    \centering
    \caption{Comparison between Biton \& Friedman~\cite{biton2025analysis} R+W and our W-only conflict graphs.}
    \label{table:read_write_vs_write_only}
    \begin{tabular}{lcc}
        \toprule
        Metric                        & Biton \& Friedman~\cite{biton2025analysis} (R+W) & This paper (W-only) \\
        \midrule
        Median density                & $0.471$                                          & $0.092$             \\
        Median assortativity          & $-0.500$                                         & $\approx+1$         \\
        Median clique number          & $5.5$                                            & $4$                 \\
        Median LCC                    & $8.5$                                            & $4$                 \\
        Median LSP/$\chi$ lower bound & $1.271$                                          & $0.667$             \\
        Median LSP/$\chi$ upper bound & $1.5$                                            & $1$                 \\
        \bottomrule
    \end{tabular}
\end{table}

By evaluating the Longest Simple Path to Chromatic Number (LSP/$\chi$) ratio across matched load regimes, we find that optimal parallel scheduling is limited under routine workloads. Biton \& Friedman~\cite{biton2025analysis} report their $3$--$5\times$ figure on the high-load 66M interval and on a synthetic $100\times$-aggregated 150Mx100 workload; on their routine 150M interval the same LSP/$\chi$ metric yields only $\approx 1.2$--$1.4\times$. Our W-only analysis gives $\approx 1\times$ at routine load and reaches up to $\approx 3\times$ at the largest checkpoints. In other words, stripping read-side edges shaves $30$--$40\%$ off the optimal-parallelism estimate at routine load and trims the high-load ceiling from $\approx 5\times$ toward $\approx 3\times$.

\begin{finding}[RQ1]
    The ``star'' structures reported in prior Sui studies are largely artefacts of shared read access to the system clock. Under a write-only model, those edges disappear, and Sui's conflict graphs are clique-dominated; the median degree-assortativity flips from $-0.5$ under R+W to $\approx+1$ under W-only.
\end{finding}

\begin{finding}[RQ2]
    At matched load regimes, the W-only estimate of achievable parallelism is $\approx 30$--$40\%$ tighter than the R+W estimate: routine-load headroom drops from $\approx 1.4\times$ to $\approx 1\times$, and the high-load ceiling drops from $\approx 5\times$ toward $\approx 3\times$.
\end{finding}

\subsection{Parallelizability Over Time via Gas-Based Analysis}

To ground these graph-theoretic limits in computational cost, we consider parallelisation potential weighted by Gas Usage, adapting methods used in related work~\cite{heimbach2023defi}.

The largest connected component (LCC) represents the largest subset of transactions that must execute sequentially, forming the upper bound of a schedule. The clique number represents the lower bound of a schedule, as it forms a complete subgraph. We define LCC as the lower bound and clique number as the upper bound of the achievable speedup in this gas-based analysis.

\begin{figure}[t]
    \centering
    \includegraphics[width=\linewidth]{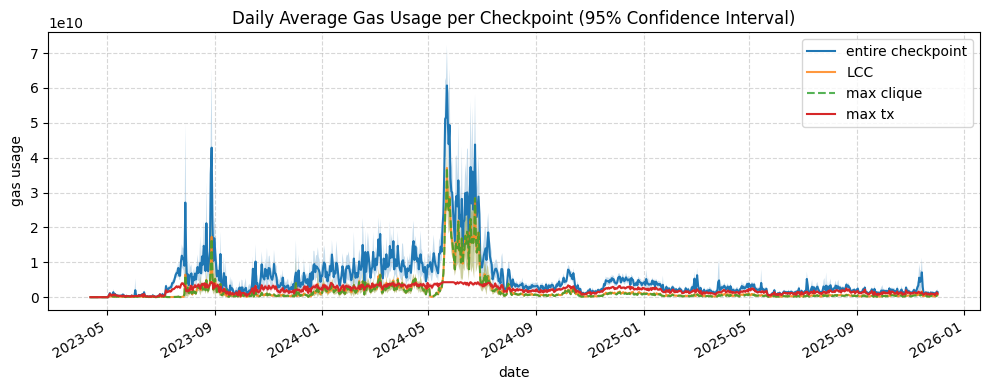}
    \caption{Historical development of gas used compared to entire checkpoint, clique number, max connected component, and max tx}
    \label{fig:daily_average_of_gas_used}
\end{figure}

As shown in \Cref{fig:daily_average_of_gas_used}, the gas used by the maximally connected component and the maximal clique frequently overlap and approach the total gas of the entire checkpoint. This indicates the presence of short-lived workloads where write-hot objects dominate the entire checkpoint, a phenomenon made possible by Sui's small checkpoint sizes.

\begin{figure}[t]
    \centering
    \includegraphics[width=\linewidth]{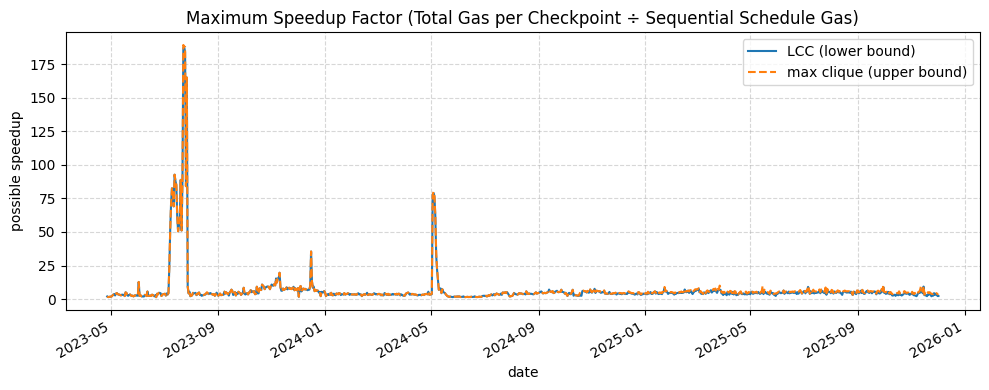}
    \caption{Lower and upper bounds of possible achievable execution speedup through parallelisation over the history of Sui}
    \label{fig:highest_speed_up_factor}
\end{figure}

\Cref{fig:highest_speed_up_factor} translates this into potential speedup factors. Consistent with our graph metrics, the possible speedup remains low and stable throughout most of Sui's history. The bounds are nearly indistinguishable. However, anomalous spikes do occur, most notably in June 2023, where potential speedup exceeded $175\times$. This spike correlates heavily with viral usage of a fully on-chain game (Sui 8192) and the Bullshark Quests initiative~\cite{Cointelegraph2023SuiTileGame}. During such periods, the massive influx of transactions interacting with a specific smart contract's shared objects temporarily shatters the typical tight bounds on parallelism.

Taken together, these two figures sharpen the count-based picture from \Cref{sec:0x6_collapse}. The gas-weighted bounds confirm that routine Sui checkpoints have effectively no residual parallelism to exploit: a small number of write-hot objects pull most of the checkpoint's gas into a single sequential chain, ruling out the ``many cheap transactions strung along a long chain'' loophole that count-based metrics could hide. The viral-period exceptions are the corollary: when the workload itself contains parallel structure, with independent users hitting disjoint objects, Sui's engine extracts very large speedups (over $175\times$ in June 2023). The low routine-load number is therefore a statement about workload composition, not about the scheduler's ability to exploit parallelism.

\begin{finding}
    The bottleneck on routine Sui load is workload composition, not the parallel engine. When the workload genuinely contains parallel structures (e.g.\ the Sui 8192 viral episode in June 2023, with achievable speedup of $\approx 175\times$), the engine extracts very large speedups; on routine load, it simply has little to parallelise. Operators should size for tail behaviour rather than average behaviour.
\end{finding}

\section{Application-Level Contention: The DeepBook Slice}
\label{sec:contention}

A central question for any parallel blockchain is whether execution contention is diffused across the network or concentrated in a few highly utilised applications. To answer this, we must first separate conflicts caused by the blockchain's internal housekeeping from conflicts driven by user activity.

\subsection{System vs. Application Contention}
\label{sec:smart_contracts}

\begin{figure}[t]
    \centering
    \includegraphics[width=\linewidth]{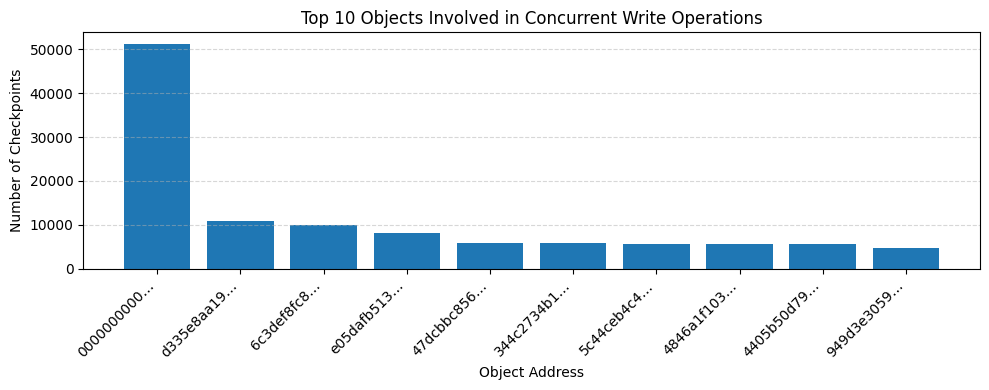}
    \caption{Top 10 objects involved in concurrent write access. The leftmost bar corresponds to the $0x6$ system clock.}
    \label{fig:top_10_objects}
\end{figure}

\begin{figure}[t]
    \centering
    \includegraphics[width=\linewidth]{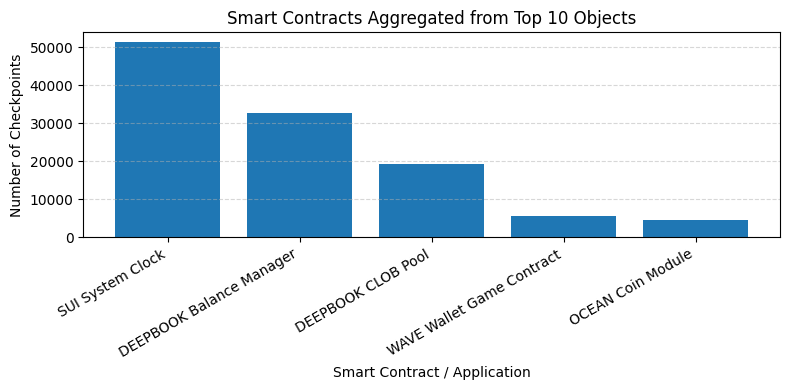}
    \caption{Smart contracts/applications associated with the top 10 objects involved in concurrent write access.}
    \label{fig:top_smart_contracts}
\end{figure}

Before analysing specific decentralised applications (dApps), we establish a baseline by identifying which objects appear most frequently in our conflict graphs. \Cref{fig:top_10_objects} lists the top $10$ objects involved in concurrent write accesses across all analysed checkpoints, and \Cref{fig:top_smart_contracts} shows the smart contracts associated with these top $10$ objects.

The most notable observation is the absolute dominance of Sui's system clock (\clock).

As discussed in \Cref{sec:0x6_collapse}, the clock is updated by system-generated prologue transactions. Our analysis reveals that the system clock is involved in more than $68\%$ of all checkpoints exhibiting conflicts.

Because only system transactions are permitted to write to the \clock object, this finding places a strict upper bound on application-level contention. It indicates that the vast majority of ``structural'' contention in the Sui network is driven by consensus mechanics and internal state management, rather than by the logic of user-facing smart contracts. User-initiated application contention therefore accounts for, at most, the remaining $32\%$ of write conflicts on the network.

Crucially, this $68\%$ figure is best understood as a measurement of Sui's \emph{design choice} to expose system state on chain via shared objects mutated by visible system transactions, rather than as a measurement of workload-driven contention. The same observation applies to the randomness beacon (\rand), the accumulator root (\acc), and any other system objects (\Cref{sec:background}). A functionally equivalent design that exposes time and randomness as validator-side metadata available to smart contracts via a built-in primitive (as Ethereum does with \texttt{block.timestamp}, or Solana with its sysvars) would produce no prologue transactions, no writes to these system objects, and no system-to-system cliques in any execution trace. Application-driven contention on the same workload would be measured identically; what would vanish is the housekeeping share that is currently visible only because Sui chose to make it visible. We therefore treat the application-only subgraph (with the system objects of \Cref{sec:background} stripped) as the cross-system-meaningful metric throughout the remainder of the paper.

\begin{finding}
    Most ($68\%$) of write-object accesses in our write-set modify a system object (the clock), with similar shares for the randomness beacon and other system objects. Under a design that handles these as validator-side metadata rather than on-chain mutable objects, this housekeeping contention would not appear in any trace; it is an artefact of Sui's specific implementation, not of the workload. Application-level contention accounts for at most the remaining $32\%$ of write conflicts.
\end{finding}

\subsection{Probing DeepBook's Role}

Having isolated the $32\%$ of the conflict graph driven by user applications, we turn our attention to the most prominent candidate for concentrated contention: DeepBook, Sui's native central limit order book (CLOB). We probe its role through two complementary experiments: a \emph{partition probe} that compares conflict structure on DeepBook-touching checkpoints against the rest, and a \emph{counterfactual probe} that simulates a perfectly parallel DeepBook by removing its internal conflict edges.

\begin{figure}[t]
    \centering
    \includegraphics[width=\linewidth]{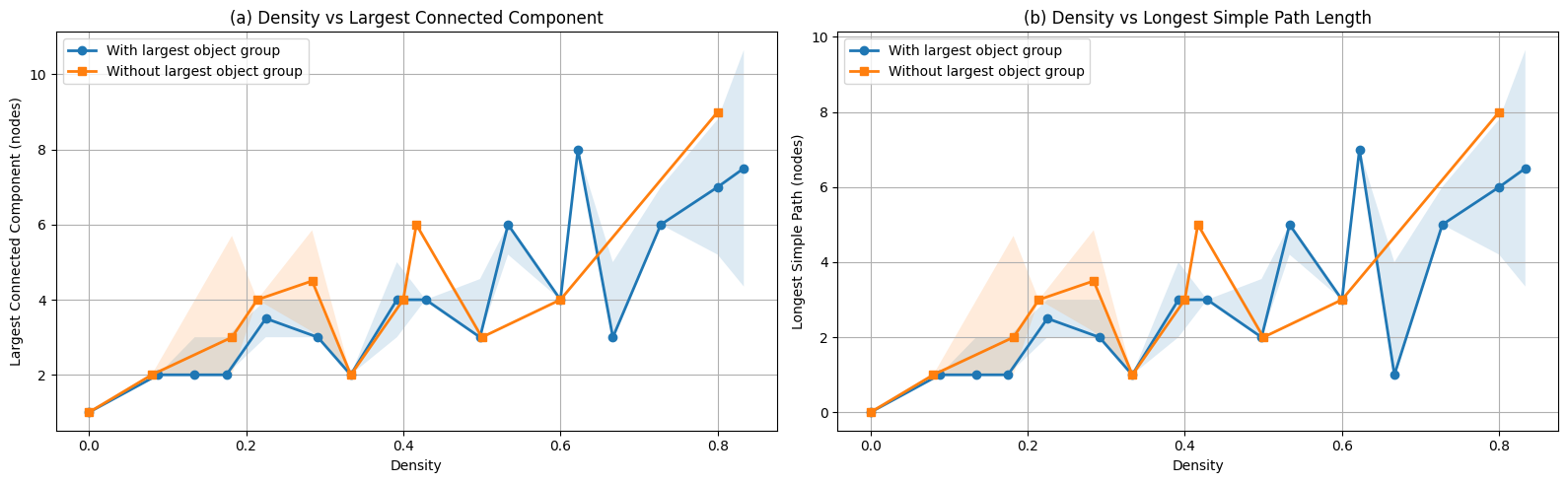}
    \caption{(a) Largest connected component (LCC) of conflict graphs, comparing checkpoints whose transactions touch the DeepBook object group against checkpoints that do not, as a function of graph density. (b) Longest simple path (LSP) of conflict graphs, comparing checkpoints whose transactions touch the DeepBook object group against checkpoints that do not, as a function of graph density.}
    \label{fig:object_groups_merged}
\end{figure}

\paragraph*{Partition probe: DeepBook vs.\ non-DeepBook checkpoints.}
To isolate the impact of DeepBook, we utilise the union-find grouping strategy detailed in \Cref{sec:methodology}. By clustering objects based on write co-occurrence, we identified a single ``giant component'' of objects that accounts for over $92\%$ of the total object write relationships on the network. By cross-referencing the object IDs in this group with known DeepBook pool IDs, we confirmed that $18$ out of $21$ active DeepBook V3 pools are contained within this single component. This confirms that our ``giant component'' effectively captures the vast majority of DeepBook's on-chain execution ecosystem.
This partitioning allows us to divide our conflict graphs into two distinct sets: checkpoints containing transactions interacting with the DeepBook ecosystem, and checkpoints that do not.
Our analysis of these partitions reveals a nuanced picture regarding sequential burden. We compared the largest connected component (LCC) and the longest simple path (LSP) across both sets. As shown in \Cref{fig:object_groups_merged}, the two sets exhibit nearly identical scaling behaviour as graph density increases: the DeepBook-touching checkpoints do not display longer sequential chains or larger connected components than the non-DeepBook ones at any given density.

This leads to a critical observation. DeepBook undeniably dominates Sui's application conflict graph by volume, participating in $87.3\%$ of all application-level conflicts. However, its contribution to the sequential bottleneck is strictly proportional to its massive transaction volume. It does not exhibit uniquely ``unparallelizable'' logic compared to the rest of the network.

This is an interesting result for a central limit order book, which is conceptually a single shared resource: in an account-based execution model, a CLOB is naturally implemented as a single contract owning both the global order book and the balance ledger, so every order placement, match, or cancellation mutates the same contract storage and forces serialisation on every operation. The pattern is well-documented in Ethereum DEX measurements~\cite{heimbach2023defi}. DeepBook's design instead exploits Sui's object model by sharding its state across separate pool objects (one per trading pair) and per-user balance manager objects, so transactions touching disjoint pools or disjoint balance managers can execute in parallel. The conflict structure we measure, contention that scales with volume but no faster than other high-volume applications, is the empirical signature of this sharding strategy. A direct head-to-head measurement against an Ethereum-style CLOB is out of scope, but the architectural argument suggests Sui's object model converts what would otherwise be a hot-contract star into a workload whose only bottleneck is volume.

\paragraph*{Counterfactual probe: removing DeepBook's internal conflicts.}
\label{sec:deepbook_edges}

To empirically verify that DeepBook's design is not an inherent bottleneck, we performed a counterfactual ``edge-removal'' experiment. In this scenario, we manually removed all conflict edges between transactions that write to the DeepBook object group. This effectively simulates a network where the DeepBook protocol is executed with ``perfect'' parallelism, free of any internal state contention.

\begin{figure}[t]
    \centering
    \includegraphics[width=\linewidth]{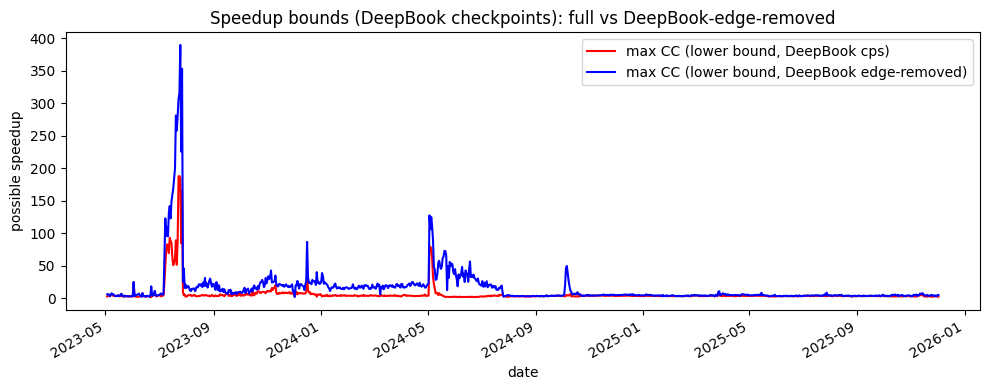}
    \caption{Potential speedup of the W-only conflict graphs with and without the DeepBook object group's internal conflict edges, across Sui's full history.}
    \label{fig:speedup_deepbook}
\end{figure}

\begin{figure}[t]
    \centering
    \includegraphics[width=\linewidth]{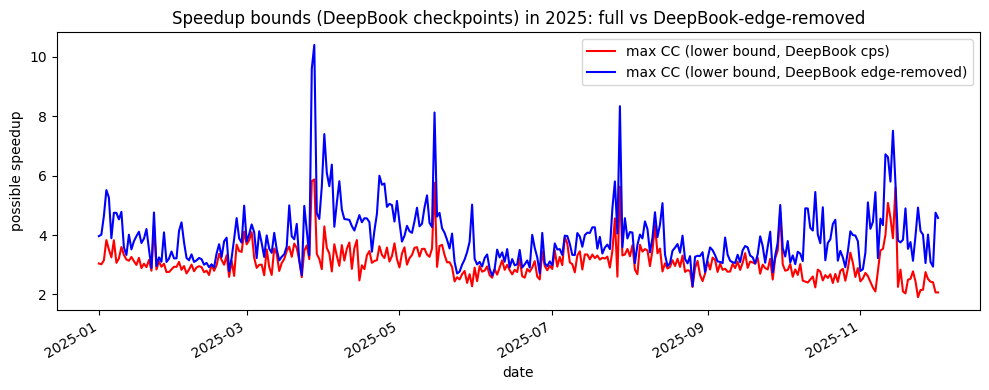}
    \caption{Potential speedup of the W-only conflict graphs with and without the DeepBook object group's internal conflict edges, restricted to 2025.}
    \label{fig:speedup_deepbook_2025}
\end{figure}

As shown in \Cref{fig:speedup_deepbook}, the resulting increase in potential speedup across the network is only moderate, typically hovering around $20\%$--$40\%$. The only exceptions are during extreme, anomalous load spikes where the entire network's capacity is stressed. \Cref{fig:speedup_deepbook_2025} restricts the same comparison to the 2025 period and shows the same pattern at a finer time resolution.
This experiment reinforces our conclusion regarding Sui's object-centric architecture. DeepBook's application state is highly distributed across many distinct objects, such as individual liquidity pools and user balances. While thousands of transactions interact with the DeepBook protocol simultaneously, they frequently touch disjoint objects, allowing Sui's execution engine to process them in parallel. The sequential chains that do inevitably form are the natural result of high transaction density in a shared market, rather than a single, poorly designed ``bottleneck'' object within the DeepBook logic itself.

\begin{finding}[RQ3]
    DeepBook accounts for $87.3\%$ of application-level conflicts on Sui by volume, but neither the partition nor the counterfactual probe shows per-transaction contention exceeding that of other high-volume applications. This is surprising for a central limit order book, normally a sequential bottleneck on account-based chains, and is explained by DeepBook's per-pool, per-balance-manager sharding of state under Sui's object model.
\end{finding}

\section{The Economic Cost of Sequentiality}
\label{sec:value}

To connect graph-theoretic contention to real-world economic flow, we measured the USD-denominated value processed in parallel versus sequential execution paths. We separate the flow of native SUI from stablecoins (USDC/USDT) to provide a dollar-denominated anchor for our analysis.

\begin{figure}[t]
    \centering

    \begin{subfigure}{0.49\linewidth}
        \centering
        \includegraphics[width=\linewidth]{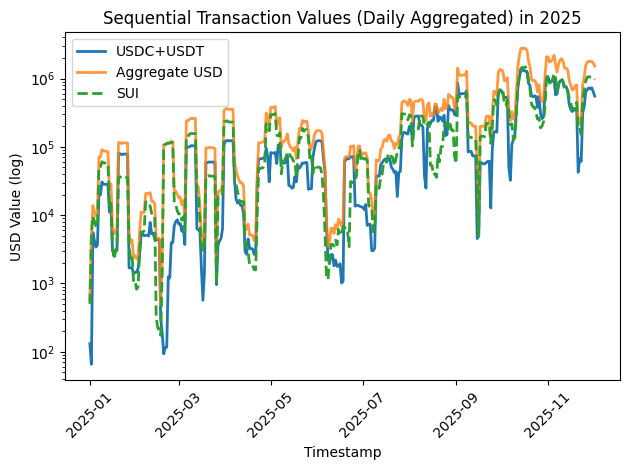}
        \caption{Transactions included in the LCC}
    \end{subfigure}
    \hfill
    \begin{subfigure}{0.49\linewidth}
        \centering
        \includegraphics[width=\linewidth]{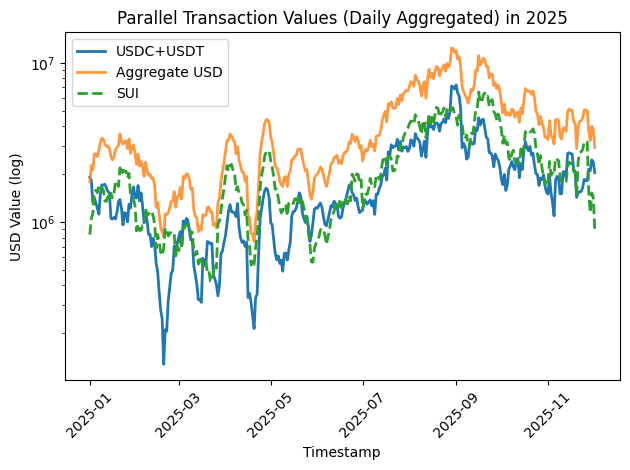}
        \caption{Transactions excluded from the LCC}
    \end{subfigure}

    \caption{Economic value of transactions included and excluded from the LCC of conflict graphs (2025).}
    \label{fig:sequential_and_parallel_transaction_values_2025}
\end{figure}

\begin{figure}[t]
    \centering
    \includegraphics[width=0.6\linewidth]{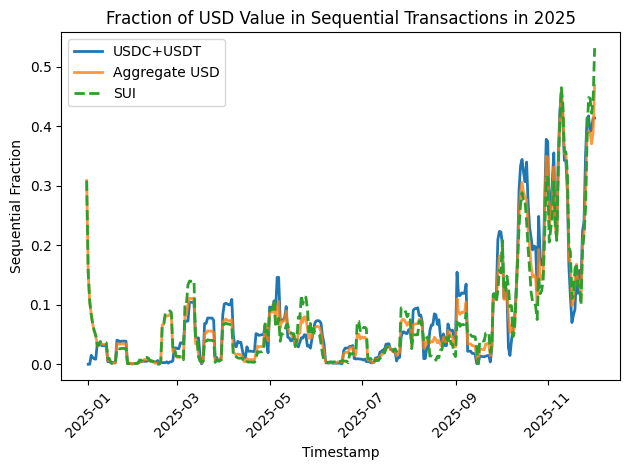}
    \caption{Fraction of transaction value subject to sequential execution (2025).}
    \label{fig:fraction_in_sequential_2025}
\end{figure}

\Cref{fig:sequential_and_parallel_transaction_values_2025} shows the economic value of sequential-burden and parallel transactions in 2025. We define ``sequential-burden'' transactions as those included in the largest connected component (LCC) of each conflict graph.
Our findings suggest that native SUI transfers dominate the measured economic activity, with a correlation coefficient of $0.999$ against the total USD-aggregated value. Interestingly, the value of parallel transactions appears more stable and larger in magnitude than that of sequential-burden transactions throughout the history of Sui.
\Cref{fig:fraction_in_sequential_2025} shows the fraction of transaction value that is subject to sequential execution over time. We find that this fraction consistently ranges between $0.1$ and $0.5$. This indicates that while Sui's parallel engine is highly effective at offloading simple transfers, a significant portion (up to 50\%) of the network's economic throughput is gated by sequential dependency chains.

Beyond describing Sui's economic flow, this sequential-burden fraction has a direct interpretation in terms of how value is currently exposed to consensus ordering. Transactions outside the LCC are parallel: their outcomes are independent of any reordering of the consensus stream. Transactions inside the LCC are ordered with respect to other LCC members on shared state, and their outcomes can, in principle, depend on that order. The $50$--$90\%$ of USD value flowing through parallel paths is therefore currently \emph{not} subject to typical Maximal Extractable Value (MEV) strategies that rely on reordering existing transactions, where MEV refers to the profit that validators can obtain by manipulating the ordering, inclusion, or exclusion of transactions within a checkpoint; the $10$--$50\%$ flowing through the sequential burden currently \emph{is potentially} subject to such strategies. This characterises only the workload as observed: a sophisticated MEV actor could submit additional transactions that introduce new conflicts and migrate some currently-parallel value into the sequential burden. Our numbers, therefore, describe the current state of the exposure of Sui's transaction value to ordering effects, not a universal MEV ceiling.

\begin{finding}[RQ4]
    In Sui's observed workload, $50$--$90\%$ of USD-denominated value flows through parallel transactions whose outcomes do not depend on consensus ordering, and is currently outside the reach of typical reordering-based MEV strategies. The remaining roughly $0$--$50\%$ flows through sequential-burden transactions whose outcomes are order-sensitive and are currently possibly affected by such strategies.
\end{finding}

\section{Related Works}
\label{sec:related_works}

This work builds on prior research on Ethereum transactions. The work of~\cite{biton2025ethereum} used the concept of read+write conflict graphs to analyse how smart contracts caused the limitations of block throughput. They found that most graphs exhibit star-like structures, and it is common to have a single transaction connected to many other transactions. By utilising the graph-based longest simple path to chromatic number ratio, the results showed that Ethereum transactions can greatly benefit from parallel smart contract execution. Another work~\cite{heimbach2023defi} analyses how decentralised finance (DeFi) and NFTs have limited concurrency of Ethereum, using gas-based analysis to show changes throughout the history of Ethereum. They showed that when transactions are disentangled, the possible speedup has increased dramatically after DeFi and NFTs were introduced. Our Sui's write-only analysis is not directly comparable to an Ethereum read+write analysis, as they have different execution models with different notions of "conflict". The work on~\cite{saraph2019empirical} studied empirical concurrency in Ethereum smart contracts and found that the speedup has steadily declined from 2016 to 2017. However, the study was not able to determine which conflicts are real and which are artefacts. Further studies on other aspects of Ethereum include, but are not limited to, Dicker et al. adding concurrency to smart contracts \cite{dickerson2017adding}, Gelashvili et al. parallel execution engine Block-STM \cite{gelashvili2023block}, and Pejic et al. analytical study of large blocks on Ethereum \cite{ocheja2024analytical}. 

Transaction execution frameworks have been widely explored in prior research. The work of~\cite{amiri2019parblockchain} proposed an order-execute paradigm that supports concurrent execution of distributed applications. Another framework~\cite{ocheja2024analytical} supports deterministic distributed transaction execution by processing an identical batch of transactions across replicas, and resolves conflicts from the reservation mechanisms prior to the execution. Unlike traditional database systems, however, blockchain environments must tolerate Byzantine failures. To address this, this work leverages distributed proposing and verification, layered consensus protocols, and signature aggregation to achieve scalable and secure transaction execution \cite{amiri2019parblockchain}. To evaluate scalability under pressure, this work presented a framework to stress-test blockchain systems using decentralised applications such as gaming and web services. While the framework provides insight into blockchain performance under application-level workloads, it does not analyse transaction conflicts or constraints on parallel execution~\cite{gramoli2023diablo}. These works primarily focus on account-centric blockchains and optimistic concurrency control (OCC) execution models, whereas Sui adopts an object-centric architecture that has incorporated parallel execution directly into its design.

Sui Lutris~\cite{sui} is a blockchain designed with sub-second finality with perpetual operation, using two approaches: Consensusless agreement and high-throughput consensus. The FastPath from the Mysticei consensus protocol \cite{mysticeti} applies uncertified direct acyclic graphs to process transactions without or before reaching consensus by using reliable broadcast to commit transactions. The Mysticeti protocol is fundamental in shaping Sui's execution path.

Sui Lutris~\cite{sui} is a high-performance blockchain designed for sub-second finality and continuous operation. Sui separates transaction processing into two execution paths: a fast path for transactions that involve only owned objects, and a consensus path for transactions accessing shared objects. The fast path leverages the Mysticeti consensus framework~\cite{mysticeti}, which uses a DAG-based reliable broadcast protocol to achieve high throughput and low latency. This separation between owned-object execution and shared-object consensus is central to Sui's object-centric execution model, which directly shapes its parallelisation behaviour analysed in this work.

Biton \& Frideman~\cite{biton2025analysis} perform a similar analysis on the Sui network using read+write conflicts, which, as mentioned previously, read+write conflicts do not capture the naive execution limits in Sui, as they capture read-only observational patterns. For example, transactions from MEV bots that watch-and-react will show up as many read edges in read+write conflict graphs. However, they are not true execution dependencies. Furthermore, the paper mentioned that the read set could be overestimated, causing the results to be a loose upper bound of possible parallelism, the set may include false dependencies, therefore, could overestimate sequentiality and underestimate achievable parallelism. This work focuses on write+write conflicts, yielding a sound, tighter bound, as the write-set contains dependencies that truly require ordering, hence, any remaining sequentiality is unavoidable. Each edge in our graph corresponds to a real, must-order dependency rather than a possibly read artefact. Any sequentiality is present and cannot be blamed on measurement noise in the read set.
\section{Limitations}
\label{sec:discussion_and_conclusion}

Our work contains several limitations.
Firstly, we selectively sampled approximately $100$ checkpoints per day in the post-Mysticeti era, and less per day in the pre-Mysticeti era. Our fixed-stride sampling method introduces temporal sampling bias between the pre and post-eras. A more rigorous approach would stratify sampling per calendar day rather than using a global checkpoint stride; Furthermore, Biton's paper~\cite{biton2025analysis} sampled the 1.5M full trace from a full node, some data may have been missed during our analysis.
Secondly, we use greedy / DFS heuristics for the NP-hard graph properties (chromatic number, clique number, longest simple path), following the standard methodology in prior conflict-graph analysis on Sui and Ethereum~\cite{biton2025analysis,biton2025ethereum}. Exact computation is intractable at the scale of our checkpoint sample; the heuristic estimates may therefore deviate from the true values, but they remain directly comparable to the bounds reported in that prior work.
Thirdly, the W-only model has two complementary blind spots relative to R+W. As described in \Cref{sec:background}, the W-only model treats \emph{write-then-read forwarding} within a single checkpoint as outside its scope: when a transaction reads a version of a shared object produced by another transaction in the same checkpoint, the reader is a \emph{candidate} execution-time consumer of the writer's output, yet the W-only conflict graph contains no edge between them. Whether the corresponding wait actually occurs depends on runtime overlap, which the consensus trace cannot resolve. Our LCC and LSP statistics may therefore under-count sequential chains in workloads with frequent intra-checkpoint forwarding, but never over-count.
Lastly, our trace-based analysis cannot precisely determine which inferred dependencies are actually constrained during runtime execution. A possible extension would be to measure the frequency of candidate intra-checkpoint write-to-read ($W \rightarrow R$) forwarding relationships using indexed \textit{object\_changes}, where a shared object written by $\mathit{tx}_i$ is later read at the new version by $\mathit{tx}_j > \mathit{tx}_i$ within the same checkpoint. Such relationships provide an upper bound on potential forwarding dependencies between transactions. However, the extent to which these dependencies truly reduce parallelism depends on runtime scheduling and execution timing information that is not captured purely in the checkpoint traces.
\section{Conclusion}

We have presented a complementary lower-bound analysis of contention in Sui based on a write-only conflict model that matches the engine's actual execution semantics. The W-only graph reveals a clique-dominated topology fundamentally different from the hub-and-spoke structures reported under read+write models, and tightens the estimate of achievable optimal parallelism by roughly $30$--$40\%$ at routine load. It also exposes that the network's largest application by transaction volume, DeepBook, contributes to contention strictly through its popularity rather than through contention-inducing smart-contract logic, a surprising property for a central limit order book that is made possible by Sui's object model and DeepBook's sharded state. Mapping contention to USD-denominated value further shows that $10$--$50\%$ of network value flows through sequential paths, bounding the share of throughput currently exposed to reordering-based MEV. Beyond Sui, the W-only methodology could extend naturally to other modern blockchains that expose explicit atomic units of state (such as Aptos's resource-based execution under Block-STM or Solana's per-account writable-flag scheduling), where it offers a cross-platform vocabulary for comparing application-driven contention without the read-side noise that dominates account-centric R+W analyses.

\bibliographystyle{plainurl}
\bibliography{references}

\appendix
\section{Database Schemas}
\label{app:schemas}

The following tables show the schemas of our PostgreSQL~\cite{PostgreSQL} indexer database. The underline represents the primary key, and the dotted line represents the foreign key.

\begin{table}[t]
    \centering
    \caption{\texttt{checkpoints}}
    \begin{tabular}{ll}
        \toprule
        \textbf{Field}               & \textbf{Type} \\
        \midrule
        \underline{sequence\_number} & Integer       \\
        digest                       & Hash Digest   \\
        timestamp                    & Timestamp     \\
        epoch\_id                    & Integer       \\
        user\_tx\_count              & Integer       \\
        system\_tx\_count            & Integer       \\
        \bottomrule
    \end{tabular}
\end{table}

\begin{table}[t]
    \centering
    \caption{\texttt{transactions}}
    \begin{tabular}{ll}
        \toprule
        \textbf{Field}                  & \textbf{Type}         \\
        \midrule
        \underline{transaction\_digest} & Hash Digest           \\
        kind                            & String                \\
        \dotuline{checkpoint\_sequence} & Integer               \\
        status                          & String                \\
        error                           & Binary-encoded String \\
        inputs\_imm\_or\_owned          & Integer               \\
        inputs\_pure                    & Integer               \\
        inputs\_receiving               & Integer               \\
        inputs\_shared\_mut             & Integer               \\
        inputs\_shared\_ro              & Integer               \\
        inputs\_funds\_withdrawal       & Integer               \\
        command\_move\_call             & Integer               \\
        command\_transfer\_objects      & Integer               \\
        command\_split\_coins           & Integer               \\
        command\_merge\_coins           & Integer               \\
        command\_publish                & Integer               \\
        command\_make\_move\_vec        & Integer               \\
        command\_upgrade                & Integer               \\
        gas\_used                       & Integer               \\
        gas\_price                      & Integer               \\
        \bottomrule
    \end{tabular}
\end{table}

\begin{table}[t]
    \centering
    \caption{\texttt{object\_changes}}
    \begin{tabular}{ll}
        \toprule
        \textbf{Field}                 & \textbf{Type}          \\
        \midrule
        \underline{object\_id}         & Integer                \\
        address                        & Binary-encoded Address \\
        \dotuline{transaction\_digest} & Hash Digest            \\
        change\_type                   & String                 \\
        input\_version                 & Integer                \\
        input\_digest                  & Hash Digest            \\
        output\_version                & Integer                \\
        output\_digest                 & Hash Digest            \\
        \bottomrule
    \end{tabular}
\end{table}

\begin{table}[t]
    \centering
    \caption{\texttt{balance\_changes}}
    \begin{tabular}{ll}
        \toprule
        \textbf{Field}                 & \textbf{Type} \\
        \midrule
        \underline{balance\_id}        & Integer       \\
        \dotuline{transaction\_digest} & Hash Digest   \\
        coin\_type                     & String        \\
        amount                         & Integer       \\
        \bottomrule
    \end{tabular}
\end{table}

We collect the write set via SQL queries from our database. The exact SQL is shown in \Cref{lst:write_set}.
\begin{lstlisting}[language=SQL, caption={SQL query used to create the WRITE set}, label={lst:write_set}]
SELECT
    t.checkpoint_sequence AS checkpoint,
    oc.transaction_digest AS tx,
    oc.address AS object_address,
    c.timestamp AS timestamp
FROM transactions t
JOIN object_changes oc
  ON oc.transaction_digest = t.transaction_digest
JOIN checkpoints c
  ON c.sequence_number = t.checkpoint_sequence
WHERE oc.change_type IS NOT NULL;
\end{lstlisting}

\section{Auxiliary Graph Metrics}
\label{app:aux_metrics}

\begin{figure}[t]
    \centering
    \includegraphics[width=\linewidth]{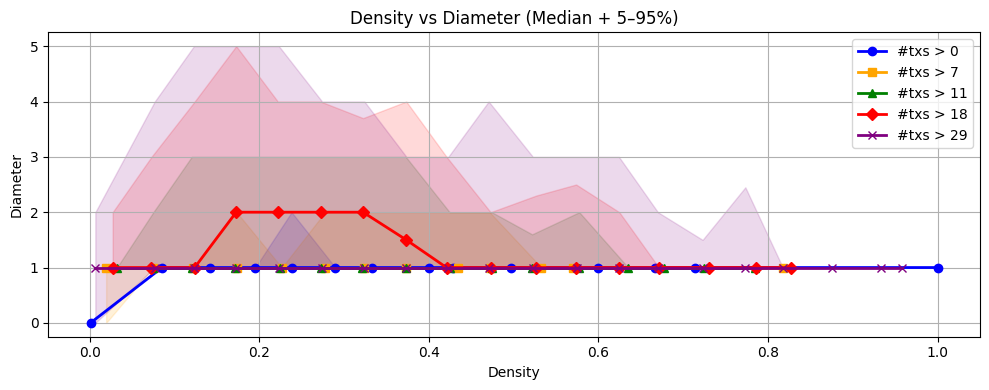}
    \caption{Concurrent access graph diameter}
    \label{fig:density_vs_diameter}
\end{figure}

In \cref{fig:density_vs_diameter}, diameter remains mostly at $1$, showing that most graphs are either cliques or tiny dense component structures. In between densities $0.2$ and $0.4$, there is an increase to $2$ for checkpoint size between $18$ and $29$, suggesting near clique graphs but slight structural imperfections, for example, a missing edge between two dense cliques.

\begin{figure}[t]
    \centering
    \includegraphics[width=\linewidth]{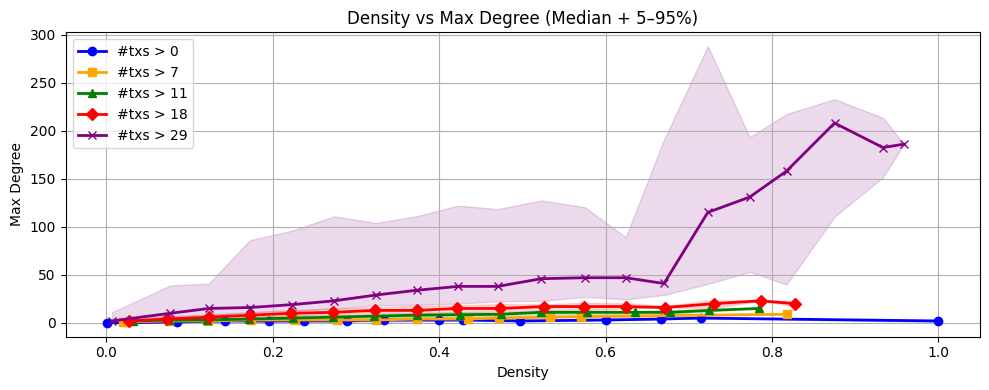}
    \caption{Concurrent access graph max degree}
    \label{fig:max_degree_vs_density}
\end{figure}

In \cref{fig:max_degree_vs_density}, the max degree of conflict graphs mostly remains small, with small linear increases as the density increases. Checkpoints from the largest quartile range show a greater, sharper increase in max degree than the rest, and it is more apparent in high densities, where it reaches up to over $200$.

\begin{figure}[t]
    \centering
    \includegraphics[width=\linewidth]{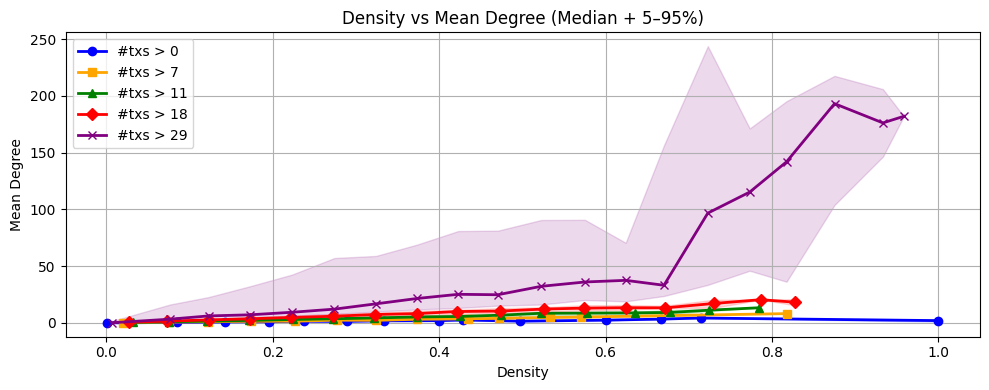}
    \caption{Concurrent access graph mean degree}
    \label{fig:mean_degree_vs_density}
\end{figure}

\Cref{fig:mean_degree_vs_density} shows the mean degree. The results are similar to the max degree, only slightly lower, as shown in the largest checkpoint range, suggesting uneven distribution. This suggests that many transactions conflict with a small number of write-hot objects.

\begin{figure}[t]
    \centering
    \includegraphics[width=\linewidth]{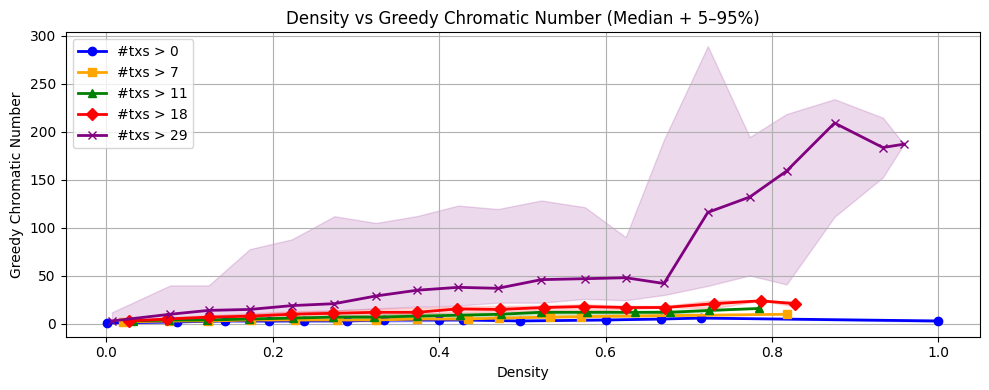}
    \caption{Concurrent access graph chromatic number}
    \label{fig:chrotamic_vs_density}
\end{figure}

\begin{figure}[t]
    \centering
    \includegraphics[width=\linewidth]{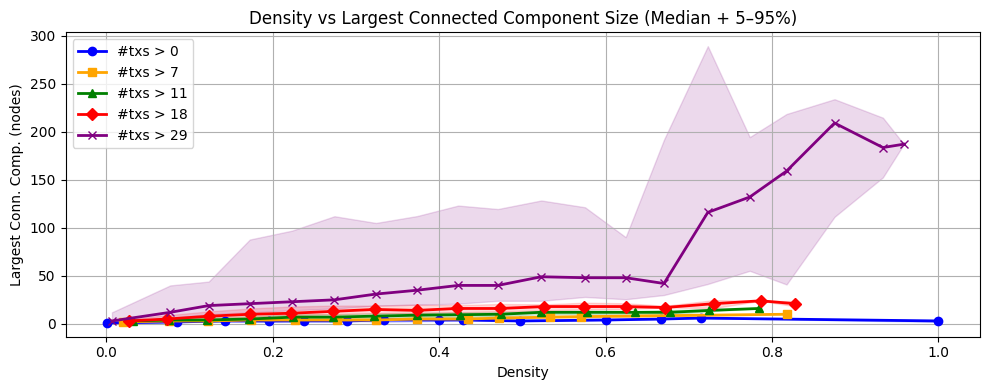}
    \caption{Concurrent access graph largest connected component by nodes}
    \label{fig:lcc_vs_density}
\end{figure}

\begin{figure}[t]
    \centering
    \includegraphics[width=\linewidth]{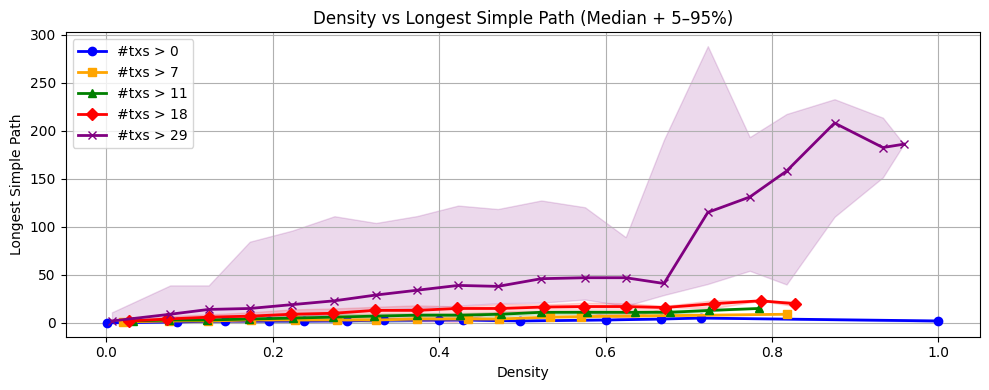}
    \caption{Concurrent access graph longest simple path}
    \label{fig:longest_simple_path_vs_density}
\end{figure}

\Cref{fig:max_degree_vs_density,fig:mean_degree_vs_density,fig:chrotamic_vs_density,fig:lcc_vs_density,fig:longest_simple_path_vs_density} show minimal variation, showing a small increase linearly as density increases, reflecting scale effects rather than structural changes in conflict graphs. The checkpoints from the largest quartile range show a greater increase, likely due to more transactions and more opportunities for overlapping object access.

\section{Input, Write, and Error Kind Breakdowns}
\label{app:kinds}

\begin{figure}[t]
    \centering
    \includegraphics[width=\linewidth]{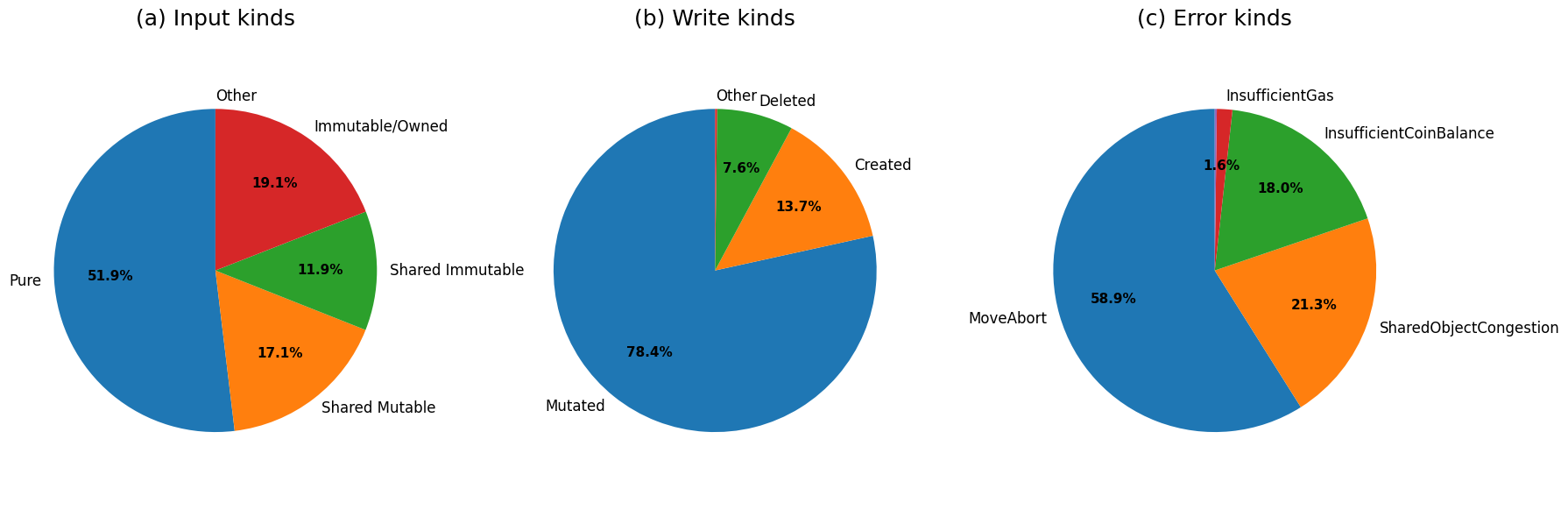}
    \caption{Partition of input kinds, write kinds, and error kinds}
    \label{fig:input_write_error_kinds}
\end{figure}

\Cref{fig:input_write_error_kinds}(a) shows the input kinds of transactions. More than half of the inputs, roughly $52\%$ are pure. Both Immutable/Owned and Shared Mutable share similar percentages, with around $19\%$ and $17\%$ respectively, while Shared Immutable makes up approximately $12\%$. Both Funds Withdrawal and Receiving inputs make up less than $0.1\%$. \Cref{fig:input_write_error_kinds}(b) shows the write kinds. Mutated dominated the chat with taking around $78\%$. This is followed by Created, with roughly $14\%$, and Deleted, around $8\%$. Others, making up from Unwrapped Then Deleted, Unwrapped and Wrapped, take up to only approximately $0.3\%$. Finally \Cref{fig:input_write_error_kinds}(c) shows the error kinds. Most errors are MoveAbort errors, with roughly $60\%$. Followed by Shared Object Congestion, with around $21\%$, and Insufficient Coin Balance, with $18\%$. There are fewer than $2\%$ Insufficient Gas errors. Other errors are insignificant, making up less than $0.2\%$ of the total. It should be noted that the analysis is post-write-only and therefore read-induced artefacts such as read-side system-clock writes are excluded.

\ifcameraready
\section{Reproducibility and Code Locations}
\label{app:reproducibility}

Our code, including the Rust + Diesel indexer and plotting notebooks, is available at:
\begin{lstlisting}[language=toml]
https://github.com/krdecade-0/sui_analysis_code
\end{lstlisting}

The analysis was built with the following Sui source-tree dependencies:
\begin{lstlisting}[language=toml]
sui-indexer-alt-framework = { git = "https://github.com/MystenLabs/sui.git", rev = "649e5e2ad4b7a48dd49fd7c0990a98f13ef5359e" }

sui-types = { git = "https://github.com/MystenLabs/sui.git",
              rev = "649e5e2ad4b7a48dd49fd7c0990a98f13ef5359e",
              package = "sui-types" }
\end{lstlisting}

CoinMarketCap dataset snapshot date: February 2026

PostgreSQL version: PostgreSQL 18
\fi

\end{document}